# Dimensional Equations of Entropy

## Amelia Carolina Sparavigna[1]

[1]Department of Applied Science and Technology, Politecnico di Torino, Italy.

**Abstract**: Entropy is a quantity which is of great importance in physics and chemistry. The concept comes out of thermodynamics, proposed by Rudolf Clausius in his analysis of Carnot cycle and linked by Ludwig Boltzmann to the number of specific ways in which a physical system may be arranged. Any physics classroom, in its task of learning physics, has therefore to face this crucial concept. As we will show in this paper, the lectures can be enriched by discussing dimensional equations linked to the entropy of some physical systems.

**1. Introduction**
In physics and engineering, dimensional analysis helps finding relationships between different physical quantities by determining some equations based on a few fundamental quantities. Usually, these fundamental quantities are length, time, mass, temperature and electric charge and are represented by symbols *L, T, M, θ* and *Q*, respectively [1]. The dimensions of any physical quantity can be expressed as products of these basic quantities, each raised to a rational power [2].

Any physically meaningful equation requires the same dimensions on its left and right sides, a property known as "dimensional homogeneity". Checking this homogeneity is a common application of dimensional analysis, routinely used to verify the plausibility of calculations. In this analysis, equations are turned into "dimensional equations".

In this paper we discuss some dimensional equations related to the entropy of some models of physics systems. In thermodynamics, entropy (usual symbol $S$) is the physical quantity linked to the second law of thermodynamics, the law which is telling that the entropy of an isolated system never decreases. This system will spontaneously proceed towards thermodynamic equilibrium, which is the configuration with maximum entropy.

Entropy is an extensive property. It has the dimension of energy divided by temperature, having a unit of joules per kelvin ($J\,K^{-1}$) in the International System of Units. However, the entropy of a substance can be given also as an intensive property, entropy per unit mass (SI unit: $J\,K^{-1}kg^{-1}$) or entropy per unit amount of substance.

The change in entropy ($\Delta S$) of a system was originally defined for a thermodynamically reversible process as $\Delta S = \int dQ_{rev}/T$, where $T$ is the absolute temperature of the system. This temperature is dividing an incremental reversible transfer of heat into that system ($dQ_{rev}$). This definition is sometimes called the macroscopic definition of entropy, because it is used without regard to any microscopic description of the thermodynamic system.

The most remarkable property of entropy is that of being a function of state. In thermodynamics, a state function is a property of the system which is depending only on the state of the system, not on the manner this system acquired that state. Several state functions exist besides entropy, all able of describing quantitatively an equilibrium state of a system. For this reason, besides the change of entropy, an absolute entropy ($S$ rather than $\Delta S$) was defined. In this case, an approach using statistical mechanics is preferred. Let us remember that the concept of entropy, which came out of thermodynamics, as proposed by Rudolf Clausius in his analysis of Carnot cycle, was linked by Ludwig Boltzmann to the number of specific ways in which a physical system may be arranged, in a statistical mechanics approach.

In fact, the modern statistical mechanics was initiated in the 1870s, by the works of Boltzmann, mainly collected and published in his 1896 Lectures on Gas Theory [3].

Entropy is then a fundamental quantity for any physics classroom, in its tasks of learning and teaching physics. As we are showing in this paper, lectures on this subject can be enriched by a discussion of dimensional equations related to the entropy of some models of physical systems. We will see some examples based on the entropy of the blackbody radiation and of the black holes, among others. Before examples, let us discuss shortly the concepts of thermodynamic and statistical entropies.

## 2. Entropy and Carnot cycle

The physical concept of entropy arose from the studies of Rudolf Clausius on the Carnot cycle [4]. This cycle is a theoretical thermodynamic reversible cycle proposed by Nicolas Léonard Sadi Carnot in 1824. It is composed of an isothermal expansion, followed by an adiabatic expansion. Then we have an isothermal compression followed by an adiabatic compression. When the Carnot cycle is represented on a pressure volume diagram (pV diagram), the isothermal stages are given by isotherm lines of the working fluid. The adiabatic stages move between isotherms. The area bounded by the cycle represents the total work done during one cycle. A Carnot cycle is also represented by using a temperature-entropy diagram (TS diagram). In such diagrams, the adiabatic reversible stage is an isentropic stage.

In a Carnot cycle, heat $Q_H$ is absorbed at temperature $T_H$ from a reservoir in the reversible isothermal expansion, and given up as heat $Q_C$ to a reservoir, with the reversible isothermal compression, at $T_C$, where $T_H > T_C$.

Through the efforts of Clausius and Lord Kelvin (William Thomson, 1st Baron Kelvin, 1824-1907), it is now known that the maximum work that a system can produce is the product of the Carnot efficiency $\eta$ and the heat absorbed from the hot reservoir:

$$W = \eta\, Q_H = \left(1 - \frac{T_C}{T_H}\right) Q_H \quad (1)$$

From the first law of thermodynamics, over the entire cycle, we have: $W = Q_H - Q_C$. Therefore, we have:

$$\frac{Q_H}{T_H} = \frac{Q_C}{T_C} \quad (2)$$

This implies that there is a function of state which is conserved over the complete Carnot cycle. Clausius called this state function as "entropy".

In the Carnot cycle, we have two entropies:

$$S_H = \frac{Q_H}{T_H}; \; S_C = \frac{Q_C}{T_C} \quad (3)$$

They have opposite signs and therefore, from (2), adding them we have zero. This results is generalized to generic reversible cycles as:

$$\oint \left(\frac{dQ}{T}\right)_{rev} = 0 \quad (4)$$

## 3. Entropy and statistical mechanics

The statistical definition was given by Ludwig Boltzmann in the 1870s. Boltzmann showed that his entropy was equivalent to that coming from Carnot cycle, within a constant number, the Boltzmann's constant.

In the Boltzmann approach, entropy is a measure of the number of ways in which a system may be arranged. This definition describes the entropy as being proportional to the natural logarithm of the number of possible microscopic configurations of the individual atoms and molecules of the system (microstates) which could give rise to the observed macroscopic state of the system.

$$S = -k \sum_i p_i \ln p_i \quad (5)$$

The sum is over all the possible microstates of the system, and $p_i$ is the probability that the system is in the i-th microstate [5]. The constant of proportionality $k$ is the Boltzmann constant. It is a constant relating energy at the individual particle level with temperature. It is also the gas constant $R$, divided by the Avogadro constant $N_A$.

The Boltzmann constant is linking macroscopic and microscopic physical quantities; for instance, for an ideal gas, the product of pressure $p$ and volume $V$ is proportional to the product of amount of substance $n$ (in moles) and absolute temperature $T$: $pV = nRT$, where $R$ is the abovementioned gas constant. Introducing the Boltzmann constant $k$ transforms the ideal gas law into an alternative form: $pV = NkT$, where $N$ is the number of molecules of gas. Therefore, the equation of the ideal gas is given in a microscopic formalism:

$$pV = nRT = NkT \quad (6)$$

For a gas, the internal energy coming from the first law of thermodynamics, is linked to the entropy by the second law, in the following equation:

$$dU = TdS - p\,dV \quad (7)$$

Let us remember that, for a fixed mass of an ideal gas, the internal energy is a function only of its temperature, whereas the entropy depends on temperature and volume.

## 4. Dimensions of entropy

The Boltzmann constant $k$ has the dimensions of energy divided by temperature, the same of entropy then. Its value in SI units is $1.380 \times 10^{-23}\, J/K$. Any dimensionless quantity, multiplied by this constant, becomes an entropy.

Let us use the symbol $E$ for the dimension [energy] and start from the Clausius entropy. The dimensional analysis gives:

$$S = \frac{dQ}{T} \rightarrow [S] = \left[\frac{E}{\theta}\right] = [k]\left[\frac{E}{E}\right]$$
$$= [k]\left[\frac{\theta}{\theta}\right] = [k]\left[\frac{k\theta}{E}\right] = [k]\left[\frac{E}{k\theta}\right] \quad (8)$$

Let us note that dimensional equations are characterized by the square brackets.

In the case we start our discussion from the statistical entropy, we have the presence of the logarithmic function too. The discussion proposed in this paper will be in the framework of Eq.8.

We have a clear example of (8) in the entropy of an ideal Fermi-Dirac gas. In a Fermi-Dirac system of particles, only one particle may occupy each non-degenerate energy level. Let us consider first the absolute zero. If there are $N$ particles, the lowest $N$ energy states are occupied up to the level $E_o$. At low temperatures, we have the entropy given by [6]:

$$S = \frac{\pi^2}{2} Nk\left(\frac{kT}{E_o}\right) \quad (9)$$

Here we can see explicitly one of the dimensional equations in (8):

$$S = \frac{\pi^2}{2} Nk\left(\frac{kT}{E_o}\right) \rightarrow [S] = [k]\left[\frac{k\theta}{E}\right] \quad (10)$$

Of course, we can also have different dimensional equations as the following:

$$[S] = [k]\left[\frac{E^\alpha}{E^\alpha}\right] = [k]\left[\frac{\theta^\alpha}{\theta^\alpha}\right] \quad (11)$$

In (11), $\alpha$ is a given power. Here, we have energies and temperature: in fact, we can find several other dimensionless ratios too, for instance of lengths, as we will see in this paper.

### 5. Entropy of the Debye model of solids
First of all, let us find an entropy which contains a ratio of temperatures.

In the Debye model, a solid is treated as an isotropic elastic continuum in which the velocity of the sound is constant. For the longitudinal and transverse waves, Peter Debye (1884-1966) put a cut-off at the upper limit of frequency to justify the fact that the solid is considered as an elastic continuum. The upper angular frequency of waves is $\omega_D$.

The model is characterized by a temperature, the Debye temperature, which is given by $\theta_D = \hbar\omega_D/k$. Let us define the dimensionless variable $x = \hbar\omega/kT$. $\hbar = h/2\pi$ is the reduced Planck constant.
The heat capacity of the solid is [7]:

$$C_V = 9Nk\left(\frac{T}{\theta_D}\right)^3 \int_0^{\theta_D/T} \frac{x^4 e^x}{(e^x-1)^2} dx \quad (12)$$

The lattice entropy is defined as:

$$S = \int_0^T C_V \frac{dT}{T} \propto Nk\left(\frac{T}{\theta_D}\right)^3 \quad (13)$$

In general:

$$S \propto Nk\left(\frac{T}{\theta_D}\right)^d \quad (14)$$

In (14), $d$ is the dimension of lattice. In the case of a two-dimensional layer $d = 2$. For a wire, $d = 1$. We can have therefore different powers as in the dimensional equation (11).

### 6. Entropy and condensed matter
Let us consider other two examples from condensed matter physics: one is concerning the vibrational entropy, the other the entropy of paramagnets.

In the previous section, we have discussed the entropy of Debye models. Of course, we can have models considering not only the elastic waves having a constant speed of the sound, but containing phonons and their true dispersions.

A phonon is a collective excitation in a periodic arrangement of atoms or molecules, such as in crystalline solids. Phonons are obtained from the second quantization of the displacement field of solids. An assembly of phonons possesses an entropy given by :

$$S = k \sum_i \left[-\ln\left(1 - e^{-\hbar\omega_i/kT}\right)\right]$$
$$+ k \sum_i \left[\frac{\hbar\omega_i/kT}{e^{\hbar\omega_i/kT} - 1}\right] \quad (15)$$

The sum is over all the frequencies and polarizations of the system [8].

Einstein proposed in 1907 that a solid could be considered an assembly of a large number of identical oscillators. All atoms oscillate with the same frequency. If we use the Einstein model of solids, and introduce the Einstein temperature:

$$\hbar\omega_E = k\theta_E \quad (16)$$

The entropy is:

$$S = 3Nk \frac{\theta_E/T}{e^{\theta_E/T} - 1} - 3Nk \ln\left(1 - e^{-\theta_E/T}\right) \quad (17a)$$

And also:

$$S = 3Nk \frac{\theta_E}{T} \frac{e^{-\theta_E/T}}{1-e^{-\theta_E/T}} \quad (17b)$$
$$- 3Nk \ln(1 - e^{-\theta_E/T})$$

Here we have a simple ratio of temperatures:

$$[S] = [k]\left[\frac{\theta_E}{\theta}\right] \quad (18)$$

The vibrational entropy of Eqs.17 appears also in the calculation of the entropy of diatomic gases. For these gases we have the contribution of rotational modes too. If the molecules have a moment of inertia $I$, at low temperatures we have that [9]:

$$S_{rot} = 3Nk \frac{\hbar^2}{IkT} e^{-\frac{\hbar^2}{IkT}}\left(1 + \frac{IkT}{\hbar^2}\right) \quad (19)$$

In (19), we have, as in (11), the following dimensional equation:

$$S_{rot} \rightarrow [S] = [k]\left[\frac{\hbar^2}{Ik\theta}\right]$$
$$= [k]\left[\frac{E^2T^2}{ML^2E}\right] = [k]\left[\frac{E^2}{E^2}\right] \quad (20)$$

Let us now discuss an example of entropy of materials having a magnetisation, in particular of a spin 1/2 paramagnet in terms of temperature and applied magnetic field [6]:

$$S = Nk \frac{1}{e^{-2\mu B/kT}+1}\ln\left(e^{-2\mu B/kT}+1\right)$$
$$+ Nk \frac{1}{e^{2\mu B/kT}+1}\ln\left(e^{2\mu B/kT}+1\right) \quad (21)$$

In (21), we have the magnetic field and the magnetic moment $\mu$. At low temperatures, the first term is negligible and then:

$$S = Nk\left(\frac{2\mu B}{kT}\right)e^{-2\mu B/kT}$$
$$[S] = [k]\left[\frac{E}{E}\right] \quad (22)$$

Here we have a ratio as in Eq.8.

### 7. Black-body radiation

A quite interesting example of dimensional equation is coming from the entropy of the black-body radiation.

Calculating the properties of radiation from a black-body was a major challenge in theoretical physics of the late nineteenth century. The problem was solved in 1901 by Max Planck in the approach which is known today as the Planck's law of black-body radiation [10].
The thermodynamics of homogeneous and isotropic electromagnetic radiation in a cavity with given volume and temperature is analysed in [11]. In this reference we find that the entropy is:

$$S = \frac{4}{3}\frac{8\pi^5 k^4}{15h^3c^3}VT^3 \quad (23)$$

Besides the Planck constant $h$, we have also the speed of light $c$.

The Planck constant is the quantum of action, introduced to describe the proportionality constant between the energy of a charged atomic oscillator in the wall of the black body, and the frequency $\nu$, of its associated electromagnetic wave. Its relevance is now fundamental for quantum mechanics, describing the relationship between energy and frequency in Planck-Einstein relation:

$$E = h\nu = \frac{h}{2\pi}2\pi\nu = \hbar\omega \quad (24)$$

In (24), we have the angular frequency $\omega$ and the reduced Planck constant $\hbar$. Action has the dimensions of [energy]·[time], and its SI unit is joule-second.
Therefore, the corresponding dimensional equation is (let us remember that in the dimensional equation $T$ means time):

$$[S] = [k]\left[\frac{k^3}{h^3c^3}L^3\theta^3\right]$$
$$= [k]\left[\frac{k^3\theta^3}{E^3}\frac{L^3}{c^3T^3}\right] = [k]\left[\frac{E^3}{E^3}\right] \quad (25)$$

### 8. An ideal Bose gas

In fact, we can write the last equation in (25), in a different form:

$$[S] = [k]\left[\frac{k^3\theta^3}{E^3}\frac{L^3}{c^3T^3}\right]$$
$$= [k]\left[\frac{L^3}{L^3}\right] \quad (26)$$

In (26), we used $[cT] = [L]$.
From (26), we have that the entropy can be the Boltzmann constant multiplied by a ratio of a given power of lengths (in this specific case, the third power). Let us discuss an example, that of an ideal Bose gas.

An ideal Bose gas is the quantum-mechanical version of a classical ideal gas which is composed of bosons. These particles have an integer value of spin and obey Bose–Einstein statistics. This statistics, developed by Satyendra Nath Bose for photons, was considered by Albert Einstein for massive particles.

In 1924 [12], Einstein deduced that an ideal gas of bosons can form a condensate at a low enough temperature. This condensate is known as the Bose–Einstein condensate. This condensate is a state of matter in which separate atoms or subatomic particles, cooled to near 0 K, coalesce into a single quantum mechanical entity, described by a wave function. As discussed in [13], there is a first-order transition, having a critical temperature $T_C$.

The entropy of this gas is given by:

$$S = \begin{cases} Nk \frac{5}{2} \frac{v}{\lambda^3} g(z) - \ln z & T > T_C \\ Nk \frac{5}{2} \frac{v}{\lambda^3} g(1) & T < T_C \end{cases} \quad (27)$$

In (27), $v = V/N$, where $V$ is the volume of the gas and $N$ the number of particles. We find also the thermal wavelength $\lambda$, which is given by:

$$\lambda = \sqrt{\frac{2\pi\hbar^2}{mkT}} \quad (28)$$

In (28), we find the mass $m$ of the particle.

The quantity $z$, which is named "fugacity", is dimensionless: $z = \lambda^3/v$ and its values are ranging from 0 to 1. The function $g(z)$ is given by:

$$g(z) = -\frac{4}{\sqrt{\pi}} \int_0^\infty dx\, x^2 \ln(1 - ze^{-x})$$
$$= \sum_{\ell=1}^\infty \frac{z^\ell}{\ell^{5/2}} \quad (29)$$

In (29), we see clearly that the entropy of the Bose gas has dimensions:

$$[S] = \left[k \frac{v}{\lambda^3}\right] = [k]\left[\frac{L^3}{L^3}\right] \quad (30)$$

**9. Bekenstein-Hawking entropy of black holes**

Can we find entropies, whose dimensional equations have a different power of lengths? The answer is positive. We have it in Ref.14, which gives the entropy of a black hole in a specific formula.

As proposed by Jacob Bekenstein, if a black hole were an object having no entropy, this fact would lead to a violation of the second law of thermodynamics [15]. In fact, when a hot gas with entropy enters a black hole, once it crosses the event horizon, its entropy would disappear. To save the second law, the black hole must be an object having an entropy, the increase of which is greater than the entropy carried by the gas. This entropy depends on the observable properties of the black hole: mass, electric charge and angular momentum. These three parameters enter only in a combination which represents the surface area of the black hole, as a consequence of the "area theorem" [16,17]. This theorem tells that the area of event horizon of a black hole cannot decrease. It is reminiscent of the law concerning the thermodynamic entropy of closed systems. As a consequence, the black hole entropy is proposed as a monotonic function of area: if $A$ stands for the surface area of a black hole (area of the event horizon), then the black hole entropy is given by:

$$S_{BH} = k\frac{A}{4L_P^2} = k\frac{c^3 A}{4G\hbar} \quad (31)$$

This entropy is known as the Bekenstein-Hawking entropy. In (31), $L_P$ stands for the Planck length:

$$L_P = \frac{G\hbar}{c^3} \quad (32)$$

while $G, \hbar, c$ denote, respectively, Newton's gravity constant, reduced Planck-Dirac constant and the speed of light.

From this equation, it is clear that:

$$[S_{BH}] = [k]\left[\frac{L^2}{L^2}\right] \quad (33)$$

In this entropy, the black hole is identified with a constant times its surface area [14]. This fact was clear after Stephen Hawking discovered that a black hole emits radiation at a well-defined temperature:

$$kT = \frac{\hbar c^3}{8\pi GM} \quad (34)$$

This temperature is also known as the Hawking radiation temperature. $M$ is the mass. The radius of a black hole is $R = 2GM/c^2$, which is the Schwarzschild radius. The surface area is then:

$$A = 4\pi R^2 = 16\pi \frac{G^2 M^2}{c^4} \quad (35)$$

The entropy is $dS = \frac{dQ}{T} = \frac{c^2 dM}{T}$ where the heat increment is identified with the energy equivalent of the in-falling mass [18]. After integration we can obtain (31).

Let us continue our dimensional analysis:

$$[S_{BH}] = \left[k\frac{A}{4L_P^2}\right] = [k]\left[\frac{c^3 A}{4G\hbar}\right]$$
$$= [k]\left[\frac{L^5 T^{-3}}{FL^2 M^{-2} FLT}\right] \quad (36)$$

Then:

$$[S_{BH}]$$
$$= [k]\left[\frac{L^5 T^{-3}}{ML^3 T^{-2} M^{-2} ML^2 T^{-1}}\right] \quad (37)$$
$$= [k]\left[\frac{L^5 T^{-3}}{L^5 T^{-3}}\right] = [k]\left[\frac{L^2 c^3}{L^2 c^3}\right]$$

In (36), $F$ means [force] and $c$ a [speed].

Let us show that, starting from the dimensions of the classical Clausius entropy, we can arrive to BH-entropy. $E$ is the energy and $W$ the work, which have the same dimensions.

$$[S] = [k]\left[\frac{k\theta}{E}\right]\left[\frac{k\theta}{W}\right] = [k]\left[\frac{k\theta}{pV}\right] \quad (38)$$

The volume $V$ is the product of area $A$ and length $L$, then:

$$[S] = [k]\left[\frac{k\theta A}{F A L}\right] = [k]\left[\frac{k\theta A L^2}{G M^2 A L}\right] \quad (39)$$

The force $F$ has the same dimension of the gravitational force. Considering that $[k\theta]$ has the same dimension of an internal energy $U$ of a perfect gas, which is dimensionally the same of a kinetic energy:

$$[S] = [k]\left[\frac{U A L^2}{G M^2 A L}\right] \quad (40)$$

Then:

$$[S] = [k]\left[\frac{M L^2 T^{-2} A L^2}{G M^2 A L}\right]$$
$$= [k]\left[\frac{L^3 T^{-3} A}{G M A T^{-1}}\right] = [k]\left[\frac{c^3 A}{G h}\right] \quad (41)$$

And these are the dimensions of Bekenstein-Hawking entropy $kc^3 A / 4G\hbar$.

Using again (35):

$$[S] = [k]\left[\frac{c^3 A}{G h}\right] = [k]\left[\frac{c^3 G^2 M^2}{G c^4 h}\right]$$
$$= [k]\left[\frac{G M^2 L^{-1}}{L^{-1} c h}\right] = [k]\left[\frac{E}{E}\right] \quad (42)$$

About the amount of entropy, in [14] an example is given. A one-solar mass Schwarzschild black hole has an horizon area of the same order as the municipal area of Atlanta or Chicago. Its entropy is about $4 \times 10^{77} k$, which is about twenty orders of magnitude larger than the thermodynamic entropy of the sun [14].

## 10. The Bekenstein bound

Bekenstein bound is the upper limit of the entropy $S$ that can be contained within a given finite region of space which has a finite amount of energy. It is also the maximum amount of information required to perfectly describe a given physical system down to the quantum level [19].

The universal form of the bound was originally found by Jacob Bekenstein as the inequality [19]:

$$S \leq \frac{2\pi k R E}{\hbar c} \quad (43)$$

where $R$ is the radius of a sphere that can enclose the given system, $E$ is the total mass–energy including any rest masses. Note that the expression for the bound does not contain the gravitational constant $G$. The Bekenstein-Hawking entropy of black holes exactly saturates the bound.

Let us consider the dimensional equation of this bound:

$$[S] = [k]\left[\frac{R E}{h c}\right] = [k]\left[\frac{L E}{h L T^{-1}}\right]$$
$$= [k]\left[\frac{E}{E T T^{-1}}\right] = [k]\left[\frac{E}{E}\right] \quad (44)$$

The bound is closely associated with black hole thermodynamics.

## 11. Entropy of vacuum

The vacuum has entropy too. In quantum mechanics and in quantum field theory, the vacuum is defined as the state of considered system with the lowest possible energy. This energy is known as the zero-point energy. Since all quantum mechanical systems undergo fluctuations, because of their wave-like nature, the vacuum is subjected to fluctuations too. The fluctuations are temporary changes in the amount of energy, as given by the Werner Heisenberg's uncertainty principle.

The zero-point energy of quantum electrodynamics was an important result in the theory of quantized fields [20,21]. The Casimir effect deals with the modification of this energy. The original analysis, proposed in [22], provided the basic problem by calculating the force between two conducting plates, due to the modification imposed by their presence, to the possible electromagnetic modes.

For an electromagnetic field, the vacuum may be considered as its equilibrium state in the limit of vanishing temperature: in [20], the authors have studied the extension to finite temperatures. Once the Casimir free energy had been calculated, by the standard thermodynamic formulae, the pressure on the plates can be obtained from it [20]. In this reference, the radiation is supposed confined between two conducting plates. The edge size of both plates is $L$. The first is placed at $z = 0$ in the XY plane. The second plate is placed at $z = a$ parallel to the XY plane. Let us suppose $L \gg a$. The entropy of the zero-point fluctuations of the fields is [20]:

$$S = k \frac{\varsigma(3)}{2^4 \pi} \frac{L^2 a}{a^3}$$
$$[S] = [k]\left[\frac{L^3}{L^3}\right] \quad (45)$$

Eq.(45) is given in the limit of high temperature. As in (26), this entropy is a ratio of cubic powers of lengths. In (45), we have:

$$\varsigma(n) = \sum_{m=1}^{\infty} \frac{1}{m^n} \quad (46)$$

In [20], it is stressed that, while the Casimir energy density vanishes in the high temperature limit, the Casimir free energy density does not. Thus, the resultant force of attraction between the plates is of entropic origin [20].

## 12. Entropic forces

Let us conclude the paper with another quite attractive problem, that of a force which, like the Casimir force, has its origin in the entropy of the corresponding physical system.

An entropic force is a phenomenological force coming for the statistical tendency of increasing entropy, rather than from a particular underlying microscopic force [23]. The first entropic approach was proposed for the Brownian motion in Ref.24.

To have a force from entropy, we can use the following dimensional equation:

$$[F] = \left[\frac{E}{L}\right] = \left[\frac{\theta S}{L}\right] = \left[\frac{\theta S L}{L^2}\right] \quad (47)$$

Let us determine the entropic force for a specific phenomenology, that of the elasticity of polymers.
Polymers can be modelled as freely jointed chains with one fixed end and one free end [25]. The length of a rigid segment of the chain is $b$; $n$ is the number of segments of length $b$. $r$ is the distance between the fixed and free ends, and $L_c$ is the contour length, equal to $nb$. When the polymer chain oscillates, distance $r$ changes over time.

The probability of finding the chain ends a distance $r$ apart is given by the following Gaussian distribution $f$:

$$f = 4\pi r^2 \left(\frac{2nb^2\pi}{3}\right)^{-\alpha} e^{-\alpha r^2/nb^2} dr \quad (48)$$

In (48), $\alpha = 3/2$. By using entropy and the Helmholtz free energy, we can obtain a force which is like that of the Hooke's law. Let us consider the entropy [25]:

$$S \approx k \ln f \quad (49)$$

The entropy is linked to the Helmholtz free energy:

$$A \approx -TS = -kT\frac{\alpha r^2}{L_c b} \quad (50)$$

And then [25]:

$$F \approx -\frac{dA}{dr} = \alpha \frac{kT}{L_c b} r \quad (51)$$

This is the law corresponding to the dimensional equation previously proposed in Eq. 47.

This example on the entropic force helps us to conclude the paper stressing the importance of dimensional analysis too. In fact, a guessed dimensional equation can suggest a new approach to solve a specific problem.